\documentclass{ws-procs9x6}
\newcommand{\beq}[1]{
\begin{equation}\label{#1}}
\newcommand{\eeq}{\end{equation}}
\newcommand{\bea}[1]{
\begin{eqnarray}\label{#1}}
\newcommand{\eea}{\end{eqnarray}}

\newcommand{\bra}[1]{\left\langle #1 \right|}
\newcommand{\ket}[1]{\left| #1 \right\rangle}

\newcommand{\Gl}[1]{Eq.~(\ref{#1})}

\newcommand{\al}{\alpha}
\newcommand{\be}{\beta}  
\newcommand{\ep}{\varepsilon}
\newcommand{\ga}{\gamma}
\newcommand{\de}{\delta}

\newcommand{\la}{\lambda}

\newcommand{\si}{\sigma}

\newcommand{\dd}{{\rm d}}

\newcommand{\nn}{\nonumber}

\begin{document}

\title{Pion-Nucleon Distribution Amplitudes}

\author{Andreas Peters}

\address{Institut f{{\"u}}r Theoretische Physik, \\
 Universit{{\"a}}t  Regensburg,\\
 D-93040 Regensburg, Germany\\
E-mail: andreas.peters@physik.uni-regensburg.de}

\begin{abstract}
This is a short presentation of the results for the pion-nucleon distribution amplitudes
which are expressed in terms of the nucleon distribution amplitudes
with the help of current algebra.
Everything is considered to be at threshold.
\end{abstract}


\bodymatter

\section{Introduction}\label{sec1}
In recent years there has been increasing attention to hard exclusive 
processes involving emission of soft pions in the final state. In the near future new
experimental data will emerge and their interpretation requires information 
about the nucleon wave function. 
Such processes are attractive as they provide us
new insights in the hadron (nucleon) structure.\\
At this point the nucleon distribution amplitudes (DAs) play an important role,
as they contain direct information about the wave function. 
Our aim will be to treat the outgoing nucleon and produced $\pi$ as a $N\pi$ final state
and describe the $N\pi$ DAs in terms of nucleon DAs. 
The main physical tool we use to calculate such $N\pi$ DAs 
is the well known soft-pion theorem. Everything is considered to be at threshold.

\section{Distribution Amplitudes (DAs)}
\subsection{Nucleon DAs of twist-3}
Let us first consider the leading
twist nucleon DAs in some detail to become acquainted with the method.
As introductory remarks we first want to repeat some notations from \cite{BFMS00}.\\
The object of interest is the
hadron-to-vacuum matrix element of a trilocal operator built of quark and 
gluon fields at light-like separations. 
\bea{dreiquark}
\bra{0} \epsilon^{ijk} u_\al^{i'}(a_1 z)\left[a_1 z, a_0 z \right]_{i',i}
u_\be^{j'}(a_2 z) \left[\ldots\right]_{j',j}
d_\ga^{k'}(a_3 z) \left[a_3 z, a_0 z \right]_{k',k}
\ket{p(P,\la)}\,,
\nn\\
\eea
where $\ket{p(P,\la)}$ denotes the proton state with 
momentum $P$, $P^2 = M^2$ and helicity $\la$. 
$u,d$ are the quark-field operators. The Greek letters $\al,\be,\ga$ 
stand for Dirac indices, the Latin letters $i,j,k$ refer to color.
$z$ is an arbitrary light-like  vector, $z^2=0$, the $a_i$ are real numbers.
The gauge-factors $\left[x,y\right]$ can be seen in \cite{BFMS00,Braun:2006td}
and keep the matrix element in (\ref{dreiquark}) gauge-invariant.
In what follows we spare on writing the gauge factors explicitely, 
but keep them always as present in mind.\\
In order to fullfill  Lorentz covariance, spin and parity conservation of the
nucleon it is convenient to write
\Gl{dreiquark} in terms of 24 invariant functions \cite{BFMS00}. In the leading twist approximation
only three amplitudes are relevant.
In a shorthand notation it reads
\bea{lt1}
&&4\bra{0} \ep^{ijk} u_\al^i(a_1 z) u_\be^j(a_2 z) d_\ga^k(a_3 z) 
\ket{p(P,\la)}_{\rm twist-3}
=
\nonumber\\&&{}= 
V^p_1 \,(v_1)_{\al\be,\ga} +  
A^p_1 \,(a_1)_{\al\be,\ga} +  
T^p_1 \,(t_1)_{\al\be,\ga}   
\eea
where 
\bea{Lor1}
 (v_1)_{\al\be,\ga}  &=&  \left(\!\not\!{p}C \right)_{\al \be} \left(\ga_5 N^+\right)_\ga 
\nonumber\\  
 (a_1)_{\al\be,\ga}  &=&  \left(\!\not\!{p}\ga_5 C \right)_{\al \be} N^+_\ga
\nonumber\\  
 (t_1)_{\al\be,\ga}  &=&  \left(i \si_{\perp p} C\right)_{\al \be} 
                            \left(\ga^\perp\ga_5 N^+\right)_\ga   
\eea
stand for the Lorentz structures. \\
The amplitudes $V^p_1$, $A^p_1$, $T^p_1$ can be written as
\beq{fourier}
F(a_i p\cdot z) = \int \! {\cal D} x\, e^{-ipz 
\sum_i x_i a_i} F(x_i)\,,
\eeq
where the functions $F(x_i)$ depend on the dimensionless
variables $x_i,\, 0 < x_i < 1, \sum_i x_i = 1$ which 
correspond to the longitudinal momentum fractions 
carried by the quarks inside the nucleon.  
The integration measure is defined as 
\beq{integration}
\int\! {\cal D} x  = \int_0^1\! \dd x_1 \dd x_2 \dd x_3\, 
\de (x_1 + x_2 + x_3 - 1)\,.
\eeq 
Applying the set of Fierz transformations
\bea{f-t3}
(v_1)_{\ga\be,\al} &=& \frac{1}{2}\left(  v_1 -  a_1 -   t_1
\right)_{\al\be, \ga} \nn \\
(a_1)_{\ga\be,\al} &=& \frac{1}{2} \left(-   v_1 +   a_1 -   t_1
\right)_{\al\be, \ga} \nn \\
(t_1)_{\ga\be,\al} &=& - \left(v_1 +  a_1 \right)_{\al\be,\ga}
\eea
one ends up with the condition
\beq{isospin2}
2 T_1^p(1,2,3) = [V^p_1-A^p_1](1,3,2) + [V^p_1-A^p_1](2,3,1)\,, 
\eeq
which allows to express the tensor DA of the leading   
twist in terms of the vector and axial vector distributions. Since the 
latter have different symmetry, they can be combined together to define 
the single independent leading twist-3 proton DA
\beq{twist-3}
\Phi^p_3(x_1,x_2,x_3) = [V^p_1 - A^p_1](x_1,x_2,x_3) 
\eeq
which is well known and received a lot of attention in the literature.  
The neutron leading twist DA $\Phi^n_3(x_1,x_2,x_3)$ can readily be obtained
by the interchange of $u$ and $d$ quarks in the defining \Gl{lt1}. 
For all invariant functions $F=V,A,T$ proton and neutron DAs differ by an overall
sign:
\beq{proneu}
    F^p(1,2,3) = - F^n(1,2,3)\,, 
\eeq
as follows from the isospin symmetry. This property is retained for all twists. 
A complete collection of the results for the higher twist DAs and the 
$x^2$-corrections can be seen in \cite{BFMS00,Braun:2006hz}.

\subsection{Pion-Nucleon DAs of twist-3}
We want to recall the three-quark DAs of a $N\pi$ pair from \cite{Braun:2006td} in the limit that the momentum of 
the pion relative to that of the nucleon is small.
The central idea is to use the well kown soft-pion 
theorem and current algebra, similar to the ideas developed in \cite{PPS01}.
To this end we 
define for the $N\pi$-system:   
\bea{lt2}
4\bra{0} \ep^{ijk} u_\al^i(a_1 z) u_\be^j(a_2 z) d_\ga^k(a_3 z) 
\ket{N(P,\la)\pi}_{\rm tw-3}&=&
\nn\\&&\hspace{-6.3cm}
(\gamma_5)_{\ga\de}\frac{-i}{f_\pi}\left[
V^{N\pi}_1 (v_1)_{\al\be,\de} +  
A^{N\pi}_1 (a_1)_{\al\be,\de} +  
T^{N\pi}_1 (t_1)_{\al\be,\de}\right].
\eea
An extra $\gamma_5$ is needed to conserve parity. Similar to the proton 
case, the symmetry of the two $u$-quarks implies that the DAs $V$ and $T$ are symmetric,
and $A$ is antisymmetric to the exchange of the first two arguments, respectively. 
Here $f_\pi=93$~MeV is the pion decay constant.\\
On the other hand one can calculate (\ref{lt2}) explicitely. 
To this end one has to evaluate (\ref{lt2}) with the help of the well known soft-pion theorem at threshold. 
For a detailed description, see \cite{Braun:2006td}. 
After some calculation one ends up with the results in leading twist \cite{Braun:2006td}:
\bea{result3}
{V}_1^{n\pi^+}(1,2,3) &=&
 \frac{1}{\sqrt{2}} \Big\{V_1^n(1,3,2)+V_1^n(1,2,3)+V_1^n(2,3,1)
\nn\\
&&+A_1^n(1,3,2)+A_1^n(2,3,1)\Big\},
\nonumber\\
{A}_1^{n\pi^+}(1,2,3) &=& 
 -\frac{1}{\sqrt{2}}       
            \Big\{V_1^n(3,2,1)-V_1^n(1,3,2)+A_1^n(2,1,3)
\nn\\
&&
+A_1^n(2,3,1)+A_1^n(3,1,2)\Big\},      
\nonumber\\
T_1^{n\pi^+}(1,2,3)&=&
\frac{1}{2\sqrt{2}}\Big\{A_1^n(2,3,1)+A_1^n(1,3,2)-V_1^n(2,3,1)-V_1^n(1,3,2)\Big\}
\nn\\
\eea
and
\bea{result2}
 {F}_1^{p\pi^0}(1,2,3) &=&\frac{1}{2}F_1^p(1,2,3)\,.
\eea 
The isospin relation similar to 
(\ref{isospin2}) is not valid any more, since the pion-nucleon pair can have both
isospin 1/2 and 3/2.   
Calculation of higher twists can be made in a complete analogy, 
but is much more extensive. 
A short review of the results is given in the Appendix.

\section{Conclusions}
We have shown the results for the $N\pi$ DAs. 
The tools we used were the light-cone formalism, 
the soft-pion theorem at threshold and current algebra.
In this way it was possible to express the $N\pi$ DAs in terms of nucleon DAs.

\section*{Acknowledgements}
I would like to thank V.~Braun, D.Yu~Ivanov and A.~Lenz 
for collaboration and G.~Peters for interesting discussion.
Special thanks are to the organizers of the {\itshape Exclusive Reactions at High Momentum Transfer 2007} 
workshop at Jefferson Laboratory for their hospitality 
and enlightening debates.
This work was supported by the Studienstiftung des deutschen Volkes.

\appendix{$n\pi^+$ and $p\pi^0$ DAs in higher twist}
 For $n\pi^+$ we obtain in twist-4:
\begin{eqnarray}
V_2^{n\pi^+}(1,2,3)&=&  -\frac{1}{\sqrt{2}}\Bigg\{
\frac{1}{2}\Bigl(\Big\{A_3^n(1,3,2)-P_1^n(1,3,2)+S_1^n(1,3,2)-T_3^n(1,3,2)
\nonumber\\
&&-T_7^n(1,3,2)-V_3^n(1,3,2)\Big\}+\Big\{1\leftrightarrow 2\Big\}\Bigr)
\Bigg\}\,,
\nonumber\\     
~
A_2^{n\pi^+}(1,2,3)&=&   -\frac{1}{\sqrt{2}}\Bigg\{
\frac{1}{2}\Bigl(\Big\{A_3^n(1,3,2)+P_1^n(1,3,2)-S_1^n(1,3,2)+T_3^n(1,3,2)
\nonumber\\&&
+T_7^n(1,3,2)-V_3^n(1,3,2)\Big\}-\Big\{1\leftrightarrow 2\Big\}\Big)
\Bigg\}\,, 
\nonumber\\     
~
V_3^{n\pi^+}(1,2,3)&=&   \frac{1}{\sqrt{2}}\Bigg\{
\frac{1}{2}\Bigl(\Big\{A_2^n(1,3,2)-P_1^n(1,3,2)+S_1^n(1,3,2)+T_3^n(1,3,2)
\nonumber\\&&
+T_7^n(1,3,2)+V_2^n(1,3,2)\Big\}+\Big\{1\leftrightarrow 2\Big\}\Bigr)
\Bigg\}\,, 
\nonumber\\
~
A_3^{n\pi^+}(1,2,3)&=&   \frac{1}{\sqrt{2}}\Bigg\{
\frac{1}{2}\Bigl(\Big\{-A_2^n(1,3,2)-P_1^n(1,3,2)+S_1^n(1,3,2)+T_3^n(1,3,2)
\nonumber\\&&
+T_7^n(1,3,2)-V_2^n(1,3,2)\Big\}-\Big\{1\leftrightarrow 2\Big\}\Bigr)
\Bigg\}\,,
\nonumber\\   
~
S_1^{n\pi^+}(1,2,3)&=&   -\frac{1}{\sqrt{2}}\Bigg\{
\frac{1}{4}\Bigl(\Big\{A_2^n(1,3,2)+A_3^n(1,3,2)+P_1^n(1,3,2)+S_1^n(1,3,2)
\nonumber\\&&
+2T_2^n(1,3,2)+T_3^n(1,3,2)-T_7^n(1,3,2)-V_2^n(1,3,2)
\nonumber\\&&
+V_3^n(1,3,2)\Big\}-\Big\{1\leftrightarrow 2\Big\}\Bigr)
\Bigg\}\,,
\nonumber\\
~
P_1^{n\pi^+}(1,2,3)&=& -\frac{1}{\sqrt{2}}\Bigg\{
\frac{1}{4}\Bigl(\Big\{-A_2^n(1,3,2)-A_3^n(1,3,2)+P_1^n(1,3,2)+S_1^n(1,3,2)
\nonumber\\&&
+2T_2^n(1,3,2)+T_3^n(1,3,2)-T_7^n(1,3,2)+V_2^n(1,3,2)
\nonumber\\&&
-V_3^n(1,3,2)\Big\}-\Big\{1\leftrightarrow 2\Big\}\Bigr)
\Bigg\}\,,
\nonumber\\
~
T_2^{n\pi^+}(1,2,3)&=& -\frac{1}{\sqrt{2}}\Bigg\{
\frac{1}{2}\Bigl(\Big\{P_1^n(1,3,2)+S_1^n(1,3,2)-T_3^n(1,3,2)+T_7^n(1,3,2)\Big\}
\nonumber\\&&
+\Big\{1\leftrightarrow 2\Big\}\Bigr)
\Bigg\}\,,
\end{eqnarray}

\begin{eqnarray}
T_3^{n\pi^+}(1,2,3)&=& \frac{1}{\sqrt{2}}\Bigg\{
\frac{1}{4}\Bigl(\Big\{A_2^n(1,3,2)-A_3^n(1,3,2)-P_1^n(1,3,2)
-S_1^n(1,3,2)\nonumber\\&&
+2T_2^n(1,3,2)-T_3^n(1,3,2)+T_7^n(1,3,2)-V_2^n(1,3,2)
\nonumber\\&&
-V_3^n(1,3,2)\Big\}+\Big\{1\leftrightarrow 2\Big\}\Bigr)
\Bigg\}\,,
\nonumber\\
~
T_7^{n\pi^+}(1,2,3)&=& \frac{1}{\sqrt{2}}\Bigg\{
\frac{1}{4}\Bigl(\Big\{A_2^n(1,3,2)-A_3^n(1,3,2)+P_1^n(1,3,2)
+S_1^n(1,3,2)\nonumber\\&&-2T_2^n(1,3,2)
+T_3^n(1,3,2)-T_7^n(1,3,2)-V_2^n(1,3,2)
\nonumber\\&&-V_3^n(1,3,2)\Big\}+\Big\{1\leftrightarrow 2\Big\}\Bigr)
\Bigg\}.
\end{eqnarray}
And similar for $p\pi^0$ in twist-4
\begin{eqnarray}
  S_1^{p\pi^0}(1,2,3)&=&-\frac{1}{2}\Big\{2P_1^p(1,3,2)-S_1^p(1,2,3)\Big\},
\nonumber\\
  P_1^{p\pi^0}(1,2,3)&=& -\frac{1}{2}\Big\{2S_1^p(1,3,2)-P_1^p(1,2,3)\Big\},
\nonumber\\
 V_2^{p\pi^0}(1,2,3) &=& \phantom{-}\frac{1}{2}V_2^{p}(1,2,3)\,, \qquad V_3^{p\pi^0}(1,2,3)=\frac{1}{2}V_3^p(1,2,3)\,,
\nonumber\\
 A_2^{p\pi^0}(1,2,3) &=& \phantom{-}\frac{1}{2}A_2^{p}(1,2,3)\,, \qquad A_3^{p\pi^0}(1,2,3)=\frac{1}{2}A_3^p(1,2,3)\,,
\nonumber\\
 T_2^{p\pi^0}(1,2,3) &=& -\frac{1}{2}T_2^p(1,2,3)\,, 
\nonumber\\
T_3^{p\pi^0}(1,2,3)&=&\frac{1}{2}\Big\{T_3^p(1,2,3)+2T_7^p(1,2,3)\Big\}\,,
\nonumber\\
 T_7^{p\pi^0}(1,2,3) &=& \frac{1}{2}\Big\{T_7^p(1,2,3)+2T_3^p(1,2,3)\Big\}.
\label{ppi0}
\end{eqnarray}
The expressions for twist-5 and twist-6 DAs are identical to those for twist-4 and twist-3, respectively, 
with some substitutions in the DAs, see \cite{Braun:2006td}.


\begin{thebibliography}{99}

\bibitem{BFMS00}
  V.~Braun, R.~J.~Fries, N.~Mahnke and E.~Stein,
  Nucl.\ Phys.\ B {\bf 589} (2000) 381
  [Erratum-ibid.\ B {\bf 607} (2001) 433].

\bibitem{Braun:2006td}
  V.~M.~Braun, D.~Y.~Ivanov, A.~Lenz and A.~Peters,
  Phys.\ Rev.\ D {\bf 75} (2007) 014021
  [arXiv:hep-ph/0611386].

\bibitem{Braun:2006hz}
  V.~M.~Braun, A.~Lenz and M.~Wittmann,
  Phys.\ Rev.\  D {\bf 73} (2006) 094019
  [arXiv:hep-ph/0604050].

 \bibitem{PPS01}
P.~V.~Pobylitsa, M.~V.~Polyakov and M.~Strikman,
Phys.\ Rev.\ Lett.\  {\bf 87} (2001) 022001.


\end{thebibliography}
\end{document}